\documentclass[12pt]{article}

\usepackage{amsmath}
\usepackage{amssymb}
\usepackage{graphicx}
\usepackage{bm}
\usepackage{upgreek}
\begin{document}

{\noindent \bf \large
Bounce universe induced by an inhomogeneous dark fluid coupled with dark matter}

\bigskip

\begin{center}

I. Brevik\footnote{iver.h.brevik@ntnu.no}

\bigskip
Department of Energy and Process Engineering, Norwegian University of Science and Technology, N-7491 Trondheim, Norway

\bigskip
V. V. Obukhov\footnote{obukhov@tspu.edu.ru}
and  A. V. Timoshkin\footnote{timoshkinAV@tspu.edu.ru}

\bigskip

Tomsk State Pedagogical University, ul. Kievskaya, 60, 634061 Tomsk and National Research Tomsk State University,  Lenin Avenue, 36, 634050 Tomsk, Russia

\bigskip

\today

\end{center}

\bigskip

\begin{abstract}
We investigate cosmological models with a linear inhomogeneous time-dependent equation of state for the dark energy, coupled with  dark matter, leading   to a  bounce cosmology. Equivalent descriptions in terms of  the equation-of-state parameters for an exponential, a power-law, or   a double exponential form of the scale factor $a$ is obtained. The stability of the solutions is explored,  by considering small perturbations around the critical points  for the bounce in the early and in the late-time universe.
\end{abstract}


\section{	Introduction}
				
Recent cosmological observations show that the current expansion of the universe is accelerating \cite{riess98}.  The current procedure to explain the cosmic acceleration consists  in  introducing a dark energy with negative pressure and negative entropy \cite{peebles03}. According to astronomical observations the dark energy currently accounts for about 73\% of the total mass/energy of the universe and only 27\% of a combination of dark matter and baryonic matter \cite{kowalski08}. The dark energy universe may have very interesting properties in the future \cite{nojiri05}. A recent  general review of dark energy is given in Ref. \cite{bamba12}. Another way to account for dark energy without any extra components is the modification of gravity  (for recent  review, see \cite{nojiri11}). Note that in such case there is possibility of unified description of early-time inflation and late-time acceleration  was first realized in the Nojiri-Odintsov model \cite{nojiri07}  of $F(R)$ gravity (for more models of unified inflation with dark energy in modified gravity, see \cite{nojiri11}). In fact, the inhomogeneous fluid cosmology under consideration may also represent some specific versions of $F(R)$ gravity \cite{nojiri07}.

In the coupled phantom/fluid model a dark energy and a dark matter are described by assuming an ideal (nonviscous) fluid with an unusual equation of state (EoS). The new forms of the equation of state for general (inhomogeneous/imperfect) dark fluid models were considered in the articles \cite{nojiri05a,nojiri06,brevik04,capozziello06}. The equation of state dark energy parameter  $w$ is known to be negative:
\begin{equation}
w=\frac{p_D}{\rho_D}<0, \label{1}
\end{equation}
where  $\rho_D$  is the dark energy and  $p_D$  is the dark pressure. According  to present observational data the value of $w$ is $w=-1.04^{+0.09}_{-0.10}$   \cite{nakamura10}. Various possible scenarios for the evolution of the universe discussed in the literature are concerned with the so-called Big Rip \cite{caldwell03,nojiri03}, the Little Rip  \cite{frampton11,brevik11,frampton12,astashenok12,astashenok12a,astashenok12b,nojiri11a,makarenko12}, the Pseudo Rip \cite{frampton12a} and the Quasi Rip \cite{wei12} cosmologies.

On the other hand in the early universe there is possibility of the matter bounce cosmological model (see Ref.  \cite{brandenberger10} and references therein).
The universe goes from an era accelerated collapse to an expanding era over the bounce without displaying a singularity as well as cyclic universe. After the bounce the universe soon
enters a phase of matter-dominated expansion. Recently the $F(R)$ and $F(T)$ gravity models in the presence of bounce cosmology was discussed in Refs. \cite{bamba13,astashenok13}. We will in the following study similar bounce cosmologies.

 We will study a cosmological system with two coupled fluids: a dark energy component with a linear inhomogeneous EoS, and a dark matter component with a linear homogeneous EoS in a flat homogeneous and isotropic FRW universe. There exist other investigations dealing with the coupling between the inhomogeneous fluid dark energy and dark matter components \cite{brevik13,timoshkin09}.
We will explore  bounce cosmological models where  the scale factor is described by exponential as well as  power-law forms and realize them in  terms of the parameters of the EoS for the dark fluid. Then we will consider the stability of these models in the early universe by considering perturbations to the first order. Also we will consider a bounce cosmological model when the  scale factor in a double exponential form. The bouncing behavior, both in the early universe and in the late-time universe,  will be presented. The stability conditions for this model are discussed.

For an introduction to bounce cosmology in general, the reader may also consult Ref.~\cite{novello08}.

\section{	Bounce cosmology via an inhomogeneous fluid}

In this section we  construct  bounce cosmological models induced by  coupled dark energy in  terms of the parameters in the equation of state.

Let us consider a universe filled with two interacting ideal fluids: a dark energy component and a dark matter component in a spatially flat Friedman-Robertson-Walker metric with  scale factor $a$. The background equations are given by [4]:

\begin{equation}
\left\{ \begin{array}{lll}
\dot{\rho}+3H(p+\rho)=-Q, \\
\dot{\rho}_m+3H(p_m+\rho_m)=Q, \\
\dot{H}=-\frac{k^2}{2}(p+\rho+p_m+\rho_m),
\end{array} \label{2}
\right.
\end{equation}				
where   $H=\dot{a}/a$ is the Hubble rate and  $k^2=8\pi G$  with $G$ denoting Newton's gravitational;  $p,\rho$  and  $p_m,\rho_m$  are the pressure and the energy density of dark energy and  dark matter correspondingly;  $Q$ is a function that accounts for the energy exchange between a dark energy and a dark matter. Here a dot denotes the derivative with respect to cosmic time $t$.
The Friedman equation for the Hubble rate is given by \cite{nojiri05}:
\begin{equation}
H^2=\frac{k^2}{3}(\rho+\rho_m). \label{3}
\end{equation}
We will investigate the cases where the scale factor has an exponential, a power-law, or a double exponential form, and study bouncing behavior around $t=0$.

\subsection{Exponential model}

Let us first assume a  bounce cosmological model where the scale factor $a$ has   an exponential form \cite{bamba13}:
\begin{equation}
a=\exp(\alpha t^2), \label{4}
\end{equation}
$\alpha$ being a positive constant. The  instant $t=0$ ($a=1$) is taken to be the instant when bouncing occurs, in accordance with earlier treatments (cf., for instance, Ref.~\cite{bamba13}).  The Hubble parameter becomes
\begin{equation}
H=2\alpha t. \label{5}
\end{equation}
We suppose that the dark energy component obeys an inhomogeneous EoS \cite{nojiri05a}:
\begin{equation}
p=w(t)\rho+\Lambda(t). \label{6}
\end{equation}
 The dark matter component is taken to obey a homogeneous EoS \cite{bamba12}:
 \begin{equation}
 p_m=\tilde{w}\rho_m. \label{7}
 \end{equation}
We will in the following assume that the dark matter is dust matter, $p_m=0$, corresponding to $\tilde{w}=0$.

Now consider the gravitational equation of motion for the dark matter:
\begin{equation}
\dot{\rho}_m+3H{\rho}_m=Q. \label{8}
\end{equation}
We choose the interaction term between  dark energy and  dark matter in the following form, depending exponentially on time:
\begin{equation}
Q(t)=Q_0\exp \left(-\frac{3}{2}Ht\right), \label{9}
\end{equation}
with $Q_0$ a constant. Thus $Q(0)=Q_0$ when  $t=0$. The solution of Eq.~(\ref{8}) becomes
\begin{equation}
\rho_m(t)=(\rho_0+Q_0t)\exp \left( -\frac{3}{2}Ht\right), \label{10}
\end{equation}
where $\rho_m(0)=\rho_0$. According to this the dark matter energy increases linearly with $t$ for low $t$, and fades away when $t\rightarrow \infty$.

Taking into account Eqs.~(\ref{2}, \ref{3}) we obtain the gravitational equation of motion
\begin{equation}
\frac{6H\dot{H}}{k^2}-\dot{\rho}_m+3H\left[ (1+w)\left(\frac{3}{k^2}H^2-\rho_m\right)+\Lambda \right]=-Q. \label{11}
\end{equation}
As the time development of the dark matter  with time is known from Eq.~(\ref{10}), Eq.~(\ref{11}) can be regarded as the equation of motion for the dark energy.

Let us choose the parameter $w(t)$ in the form
\begin{equation}
w(t)=-1-\frac{\delta k^2}{3}H^2, \label{12}
\end{equation}
 with $\delta$ a positive constant. Thus $w(t)<0$ always, corresponding to the phantom region. When $t\rightarrow 0$, $w(t)\rightarrow -1$ (the cosmological constant case), while when $t\rightarrow \infty$, $w(t)\rightarrow -\infty$.

 Solving Eq.~(\ref{11}) with respect to $\Lambda(t)$ (the time-dependent cosmological 'constant') when $w(t)$ is taken to have the form (\ref{12}),  we obtain
 \begin{equation}
 \Lambda(t)= -\frac{4\alpha}{k^2}+\delta H^4-\left(w+\frac{3}{2}\right) \rho_m. \label{13}
 \end{equation}
Thus  when $t\rightarrow 0$ one sees that $\Lambda \rightarrow -4\alpha/k^2-\rho_0/2$ (cf. Eq.~(\ref{10})), while when $t\rightarrow \infty$, $\Lambda(t)\rightarrow \delta (2\alpha t)^4 \rightarrow \infty$.

 Concluding this subsection, we have  explored the exponential model of bounce cosmology in terms of time-dependent parameters in the EoS, taking into account the interaction between dark energy and dark matter.

 \subsection{Power-law model}

 We next consider the case where the scale factor has the following form \cite{bamba13} (see also \cite{timoshkin09}):
 \begin{equation}
 a=\bar{a}\left(\frac{t}{\bar{t}}\right)^q+1, \label{14}
 \end{equation}
 where $\bar{a} \neq 0$ is a constant, $\bar{t}$ is a reference time, and $q=2n, n \in Z$.

 In this case the Hubble parameter becomes
 \begin{equation}
 H=\frac{\bar{a}(q/\bar{t})( t/\bar{t})^{q-1}}{\bar{a}(t/\bar{t})^q+1}. \label{15}
 \end{equation}
 In analogy to the case above, we assume also now that the dark matter pressure is zero, $p_m=0$.

 The derivative of  $H$ with respect to cosmic time   is equal to
 \begin{equation}
 \dot{H}=H\left[ \frac{t^{-1}(q-1)}{\bar{a}(t/\bar{t})^q+1}-\frac{H}{q}\right]. \label{16}
 \end{equation}
 Let us choose the interaction term between dark energy and dark matter in the form
 \begin{equation}
 Q=\frac{Q_0}{\left[\bar{a}(t/\bar{t})^q+1\right]^3}. \label{17}
 \end{equation}
 As before, $t=0$ means the bouncing time, corresponding to $Q(t=0)=Q_0$. When $t\rightarrow \infty$, the interaction term $Q(t)\rightarrow 0$.

 Solving the gravitational equation of motion (\ref{8}) for the dark matter, we find
 \begin{equation}
 \rho_m(t)=\frac{\rho_0+Q_0t}{\left[ \bar{a}(t/\bar{t})^q+1\right]^3}, \label{18}
 \end{equation}
 where $\rho_m(0)=\rho_0$.

 Let us assume that the thermodynamic parameter  $w(t)$ in the EoS (\ref{6}) has the same form (\ref{12}) as before. Taking into account Eqs.~(\ref{15})-(\ref{18}) we obtain from Eq.~(\ref{11}) the following expression for the cosmological 'constant':
 \begin{equation}
 \Lambda(t)=\delta H^4-\frac{2H}{k^2}\left[ \frac{t^{-1}(q-1)}{\bar{a}(t/\bar{t})^q+1}-\frac{H}{q}\right] -\left[w-\bar{a}\left(\frac{t}{\bar{t}}\right)^q\right]\rho_m. \label{19}
 \end{equation}
Near $t=0$ for $n\neq 1$  we  have $\Lambda \rightarrow \rho_0$. In the case $n=1$ we have $\Lambda \rightarrow -4\bar{a}/k^2\bar{t}^2+\rho_0$.

As a result we have explored how the power-law expression (\ref{14}) for the scale factor, in addition to the form (\ref{17}) for $Q$ and the form (\ref{12}) for $w(t)$,  influence the  parameter  $\Lambda(t)$ in the EoS for the bouncing universe.

\subsection{Double exponential model}

 Now we will consider a double exponential model where the scale factor has the form \cite{bamba13}	
 \begin{equation}
 a=\exp(Y)+\exp(Y^2), \label{20}
 \end{equation}				
  where   $Y=(t/\bar{t})^2$ and  $\bar{t}$  is as before a reference time. In this model there is a unification of the bouncing behavior both in the early universe and in the late-time cosmic acceleration.

The Hubble parameter is equal to
\begin{equation}
H=\frac{2}{\bar{t}}\sqrt{Y}\,\frac{\exp(Y)+2Y\exp(Y^2)}{\exp(Y)+\exp(Y^2)}. \label{21}
\end{equation}
We suppose that the interaction term $Q$ has the form
\begin{equation}
Q=\frac{Q_0}{a^3}, \label{22}
\end{equation}
with $Q=Q_0$ at $t=0$.

The we can find the solution of the gravitational equation of motion for the dark matter \cite{nojiri05a}:
\begin{equation}
\rho_m(t)=\frac{\rho_0}{a^3}+Q\bar{t}\,\sqrt{Y}, \label{23}
\end{equation}
from which
\begin{equation}
\dot{\rho}_m(t)=Q-3H\rho_m(t). \label{24}
\end{equation}
If we take the parameter $w(t)$ to have the same form (\ref{12}) as before, we can solve $\Lambda(t)$ from Eq.~(\ref{11}):
\begin{equation}
\Lambda(t)=\delta H^4-\frac{2}{k^2}\left\{ \frac{H}{\bar{t}\sqrt{Y}}
-H^2+\left( \frac{2}{\bar{t}}\right)^2Y\left[ 1+\frac{(1+4Y^2)\exp(Y^2)}{a}\right] \right\}+w\rho_m. \label{25}
\end{equation}
When  $t\rightarrow 0$ the cosmological 'constant' $\Lambda \rightarrow -2/(k^2\bar{t})-\rho_0$.

As result we have constructed the double exponential model in the terms of the EoS parameters  for the coupled dark energy.

\section{Stability of the solutions}
			
In this section we  examine the stability of the solutions obtained in the previous section. First of all, we establish the formalism of this method. The evolution equations (\ref{2}) can be expressed as a plane autonomous system \cite{nojiri05}
\begin{equation}
\left\{ \begin{array}{lll}
\frac{dx}{dN}=-[1+p'(\rho)]\left[3+\frac{Q}{2H\rho_k}\right]x+3x[2x+(1+w_m)(1-x-y)], \\
\frac{dy}{dN}=-[1-p'(\rho)]\left[3x+\frac{Q}{2H\rho_v}y\right]+3y[2x+(1+w_m)(1-x-y)], \\
\frac{1}{H}\frac{dH}{dN}=-\left[3x+\frac{3}{2}(1+w_m)(1-x-y)\right].
\end{array} \label{26}
\right.
\end{equation}
here we have defined nondimensional quantities,
\begin{equation}
x=\frac{k^2\rho_k}{3H^2}, \quad y=\frac{k^2\rho_v}{3H^2}, \label{27}
\end{equation}
where $\rho_k$ and $\rho_v$ are the "kinetic" and "potential" terms,
\begin{equation}
\rho_k=\frac{1}{2}(\rho+p), \quad \rho_v=\frac{1}{2}(\rho-p), \label{28}
\end{equation}
and $N=\ln a$, $p'(\rho)=dp/d\rho$.

We shall investigate the stability of the critical points in the bounce models. Let us consider the early universe ($t\rightarrow 0$), where bounce cosmology can be realized. We assume for simplicity that the bounce takes place at $t=t_*$, so that $t*<0$ $(t_*>0)$ represents the contracting (expanding) phase. The bounce time $t_*$ is assumed to be very small. In this model the universe has a collapsing era for $t<t_*$, and thereafter initiates an expanding era.

With these assumptions the autonomous system (\ref{26}) for the exponential, the power-law, and the double exponential models can be written on the common form
\begin{equation}
\left\{ \begin{array}{ll}
\frac{dx}{dN}=3x(1+x-y), \\
\frac{dy}{dN}=3[y(1+x-y)-2x].
\end{array} \label{29}
\right.
\end{equation}
Setting  $dx/dN=0$ and $dy/dN=0$ in Eqs.~(\ref{29}), one obtains for the  critical points: $(x_c', y_c')= (0,0)$  and  $(x_c'', y_c'')=(0,1).$

Consider now small perturbations $u$ and $v$ around the critical point $(x_c.y_c)$, i.e.,
\begin{equation}
x=x_c+u, \quad y=y_c+v. \label{30}
\end{equation}
With $(x_c',y_c')=(0,0)$ the substitution of (\ref{30}) into (\ref{29}) leads in the linear approximation to the first order differential equations
\begin{equation}
\left\{ \begin{array}{ll}
\frac{du}{dN}=3u, \\
\frac{dv}{dN}=3(v-2u).
\end{array} \label{31}
\right.
\end{equation}
The general solution of (\ref{31}) for the evolution of linear perturbations can be written as
\begin{equation}
\left\{ \begin{array}{ll}
u=c_1\exp(3N), \\
v=(c_2+c_3N)\exp(3N),
\end{array} \label{32}
\right.
\end{equation}
where $c_1,c_2$ and $c_3$ are arbitrary constants.

In the limit $N\rightarrow \infty$, $u,v \rightarrow \infty$. There occurs an asymptotic instability of the solution around this critical point. This instability may be a consequence of of the influence from dark matter on dark energy.

 In the case  $(x_c'',y_c'')=(0,1)$ the autonomous system (\ref{29}) in the linear approximation becomes
 \begin{equation}
 \left\{ \begin{array}{ll}
 \frac{du}{dN}=0, \\
 \frac{dv}{dN}=-3(u+v).
 \end{array} \label{33}
 \right.
 \end{equation}
The solution of (\ref{33}) has the form
\begin{equation}
\left\{ \begin{array}{ll}
u=c_1, \\
v=c_2+c_3\exp(-3N).
\end{array} \label{34}
\right.
\end{equation}
In the case $c_3=0$ we obtain a one-parametric family of points at rest on a straight line. The resting point $(0,1)$ is stable.

This concludes our study of stability of the solutions of the bounce cosmological models.

\section{	Conclusion	}
						
In the present paper we  investigated  bounce cosmological models in which we  took into account the  interaction between  dark energy  and dark matter.   In terms of  parameters of the equation of state, $w(t)$ and $\Lambda(t)$ (cf. Eq.~(\ref{6})), we  described  bounce cosmologies when the scale factor is expressed by either an exponential, a power-law, or a  double exponential form. In all cases, we  adopted the form (\ref{12}) for the parameter $w(t)$. It turned out that near the bouncing instant $t=0$, the expressions for $\Lambda(t)$ reduced essentially to the initial density of dark matter, $\rho_0$.

Not very much is known from observations about the form of the interaction term $Q$. The analytic forms given above, in Eqs.~(\ref{9}), (\ref{17}), and (\ref{22}), were motivated chiefly by mathematical tractability. Physically, we have as checking points that $Q\rightarrow 0$ when $t\rightarrow \infty$.

 We also  analyzed in a linear  approximation the stability of the stationary points against perturbations and  showed the existence of a stable point, and attractor solutions, for these models.

\bigskip

{\bf Acknowledgments}

\bigskip

This  work was  supported by the grant of  Russian Ministry of Education and Science, project TSPU-139 (AVT and VVO).

\end{document}